# How the coupling of green finance and green technology innovation affect synergistic effect of pollution and emission carbon reduction: evidence from China

Guoqiang Liu, Ruijun Xie*
School of Statistics and Applied Mathematics, Anhui University of Finance and Economics,
Bengbu 233030, China;
*Correspondence: xieruijun@aufe.edu.cn

**Abstract:**
Amid China's "dual-carbon" transition, the synergistic alignment of green finance with green-technology innovation is pivotal for co-controlling pollution and $CO_2$ emissions. Using panel data for 266 Chinese prefecture-level cities over 2007–2023, We construct the coupling coordination index system of green finance and green technology innovation via a coupling-coordination model and systematically analyzes influencing mechanism of synergistic effect of pollution and carbon reduction. Four findings emerge. (1) The coupled-coordination significantly enhances the synergy, and energy efficiency plays a partial intermediary role in the relationship between the two. (2) The effect is heterogeneous: pronounced in the eastern and western regions, negligible in the central region, and stronger in non-resource-based and non-Yangtze River Basin cities. (3) A double-threshold model reveals a non-linear strengthening pattern as green-finance depth increases. (4) Spatial Durbin estimates show positive spillovers: the coupling of green finance and green technology innovation not only improves the level of local coordination, but also drives the improvement of environmental performance in adjacent areas. These results provide quantitative guidance for allocating green-finance resources, elevating green-innovation efficiency, and designing regionally coordinated mitigation policies.

## 1. Introduction

With the intensification of global climate change, China's environmental governance has stepped into a critical period. The "synergy of pollution reduction and carbon reduction" has become the core path to achieve the goal of "double carbon". The report of the 20th National Congress of the Communist Party of China clearly proposed to improve the green fiscal, taxation, financial and investment policy system, and put green finance and green technology innovation at the hub of green development strategy. The real contradiction is that traditional high energy consuming enterprises are facing the transformation bottleneck of "difficult financing and expensive research and development", and green technology projects are frequently lacking due to "long cycle and high risk", which urgently needs the "coupling interaction" between green finance and green technology innovation, that is, finance provides capital and risk sharing for technology, technology enhances financial sustainability with high return and verifiable environmental benefits, and promotes the governance mode from "passive governance at the end" to "active transformation at the source".

However, the coordinated development of green finance and green technology is still subject to three constraints: first, the regional gradient difference is significant, the "double high" coupling of Finance and technology in the East is close, and the "double low" in the west is difficult to form an effective investment innovation closed loop, resulting in the "strong east and weak West" effect of pollution reduction and carbon reduction; Second, the mechanism threshold effect is prominent. In the initial stage, due to asymmetric information, inconsistent evaluation standards and long transformation cycle, green finance's support efficiency for green technology innovation is low. Only when the market scale, evaluation system and risk mitigation tools become mature, the incentive effect will be fully released; Third, the channel of space spillover is not smooth, and the lack of cross regional resource allocation and technology sharing platform of green finance has restrained the diffusion of advanced green technology to less developed regions and blocked the nationwide spread of synergy. In this context, it is of great practical urgency to explore how the coupling mechanism of green finance and green technology innovation affects the synergistic effect of pollution reduction and carbon reduction, and clarify the path of the two in different development stages and regions, so as to optimize the allocation of green financial resources, improve the efficiency of green technology innovation, and promote the realization of the "double carbon" goal.

As an important financial model to promote sustainable development, the connotation of green finance is gradually clear in academic research and policy practice. As for the definition of green finance, it is adopted by many research institutes in the guidance on building a green financial system issued by seven ministries and commissions including the people's Bank of China in 2016, which emphasizes that green finance realizes "the coordination of environmental objectives and financial functions" through financial instrument innovation. Scholars summarize it as "an environment-friendly and sustainable development oriented financial model", and the core is to guide social capital to flow into low-carbon fields by optimizing the allocation of financial resources[1,2].Some scholars also quoted the definition of G20 green finance research group, emphasizing that green finance is "an investment and financing activity that can produce environmental effects and support sustainable development"[3].In the research of green finance, scholars mostly construct policy dummy variables to investigate the impact on enterprise pollution reduction[4], enterprise green technology innovation[5], regional green development[6], green innovation agglomeration[7] and other aspects.In addition, some scholars measured green finance from the four dimensions of green credit, green securities, green investment and green insurance[8-10], while others increased carbon finance[11], emphasizing that it guided the flow of funds to resource-saving and eco-friendly industries through differentiated lending policies, so as to realize the "reallocation of funds from high pollution fields to green fields".

Green technology innovation, as the core power to achieve ecological and environmental protection and sustainable economic development, focuses on the "synergy of technological innovation and environmental benefits". Some scholars believe that green technology innovation needs to incorporate environmental factors into the whole process of innovation, and the core is to reduce the negative environmental externalities of the production process through technological improvement[12,13].Some scholars also defined it as "technological innovation that takes the ecological economy as the principle, realizes pollution reduction and emission reduction, and weakens the negative externality of enterprise development on the ecological environment", emphasizing its practical effect in reducing pollutant emissions and improving energy efficiency[14].Many scholars' research on green technology innovation covers a wide range of perspectives, including the perspective of financial support[15], environmental protection target responsibility system[16], collaborative governance[17], new productivity theory[18], industrial upgrading[19], digital governance[20], etc., to analyze the impact of green technology innovation from multiple perspectives. From a macro perspective, scholars focus on the impact of industrial and financial cooperation on urban low-carbon transformation with heterogeneous green technology innovation research[21], and discuss its impact on urban green technology innovation with emissions trading scheme policy impact [22]and high-tech industry agglomeration[23]. From the micro perspective, scholars study and analyze the driving mechanism of environmental regulation punishment on enterprise environmental information disclosure through green technology innovation[24]. Part of the research will focus on the impact of capital market factors[25], trade liberalization[26] and the position of enterprises embedded in the global value chain on green innovation[27].

The research on the synergistic effect of reducing pollution and carbon has formed a multi-scale system of "macro Region - meso city - micro enterprise", and presented a logical chain of "mechanism - Measurement - mechanism - policy". In regional studies, scholars have found that land use change is the main cause of spatial heterogeneity, exposing the neglect of current governance on ecosystem service value[28]. From the perspective of cities, the "community effect" is fragmented[29], the degree of urban synergy in China is "high in the East and low in the west"[30], and the hot spots are evolving from coastal areas to inland areas. It is found that energy structure optimization and green process innovation are the key intermediaries in enterprises[31], and there is a "symbiotic benefit" and "conflict relationship" in industrial processes[32]. From the perspective of technical elements, it is explored that environmental regulation has positive regulation[33]. Digital technology innovation drives synergy through optimizing energy structure, industrial upgrading and digital intelligence integration. It has stronger effect in resource-based and key pollution control cities[34]. Its spatial correlation network is "Eastern core - Central and western edge". Network connectivity reduces regional synergy costs, but administrative barriers weaken spillovers[35]. The new infrastructure improves collaborative performance by alleviating the mismatch of innovation factors. Technology intensive cities benefit more, but it is necessary to prevent the rebound of high energy consumption in infrastructure[36].

Research on green finance, green technology innovation and pollution and carbon reduction has accumulated, but the coupling analysis of the three is still weak. The research on green finance focuses on the effect of instruments, ignoring the dynamic interaction with green technology innovation and the "capital technology" closed-loop mechanism; The research on green technology innovation focuses on the influencing factors, and lacks the discussion on feeding back the financial system; Most of the researches on reducing pollution and carbon regard them as independent variables, and their coupling synergy effect is not quantified. In addition, research methods rely on linear models, which are difficult to reveal the characteristics of nonlinearity, spatial spillover and multi factor linkage, which restricts the in-depth understanding of system complexity and regional and industrial heterogeneity. Based on this, taking 266 prefecture level cities from 2007 to 2023 as samples, this paper systematically investigated the mechanism and threshold effect of the coupling and coordination of green finance and green technology innovation on the synergistic effect of pollution and carbon reduction. This study not only helps to enrich the theoretical framework of the coupling coordination of green finance and green technology innovation and the collaborative governance of pollution reduction and carbon reduction, but also provides practical guidance for cities to achieve green and low-carbon transformation.

## 2. Theoretical Analysis and Research Hypotheses

*2.1. Influence mechanism of coupling coordination between green finance and green technology innovation on the synergistic effect of pollution and carbon emission reduction.*

Green finance promotes the optimization of industrial structure and reduces pollution and carbon by guiding capital flow to green industries. When making investment decisions, green financial institutions tend to choose projects and enterprises that are environmentally friendly and low-carbon[37].The guiding role of such funds promotes more social resources to flow into green industries, and promotes the transformation of traditional high pollution and high energy consumption industries to green and low-carbon industries[38].Green technology innovation provides technical support for this transformation, which enables the optimization of industrial structure to be smoothly realized, and then realizes the synergistic effect of reducing pollution and carbon. Green technology innovation can directly improve resource utilization efficiency and reduce pollutants and carbon emissions. With the support of green finance, enterprises have more capital and power to carry out green technology research and application[39]. For example, through the research and development of efficient energy conversion technology, improve the efficiency of energy utilization, so as to reduce energy consumption and corresponding carbon emissions; The development of new pollution control technologies can directly reduce the emission concentration and quantity of pollutants. The coupling of green finance and green technology innovation has accelerated the research, development and application of these technologies, making the effect of reducing pollution and carbon more significant. Based on the above analysis, the following hypothesis was proposed:

**Hypothesis 1.** The improvement of the coupling and coordination between green finance and green technology innovation helps to develop the synergistic effect of reducing pollution and emission carbon.

*2.2. Coupling coordination between green finance and green technology innovation and energy efficiency*

Energy transformation is an important way to achieve the "double carbon" goal and national high-quality development. Green finance and green technology innovation can provide financial and technical support for enterprises to adopt more efficient energy utilization technologies and improve energy efficiency. For example, green financial institutions provide low interest loans to enterprises that develop energy-saving equipment, prompting enterprises to increase R&D investment, develop equipment with better energy-saving effect, and then reduce the energy consumption per unit output, that is, improve energy efficiency[40].Green finance can explore new ways of coordination and cooperation between the financial sector and local governments to help regional green development[41]. The achievements of green technology innovation, such as new energy-saving technology and efficient energy conversion technology, can be directly applied to the production process, reduce energy waste and improve energy efficiency[42]. The improvement of energy efficiency means that less energy is consumed in the production process to achieve the same output, which helps to reduce pollutants and carbon emissions[43]. Because the exploitation, processing and use of energy are the main sources of pollutants and carbon emissions, the higher the energy efficiency, the less the pressure on the environment, which is conducive to the improvement of the collaborative level of pollution and carbon reduction.Based on the above analysis, the following hypothesis was proposed:

**Hypothesis 2.** The coupling coordination of green finance and green technology innovation promotes pollution reduction and carbon reduction by improving energy efficiency.

*2.3. Coupling coordination between green finance and green technology innovation and the level of financial development*

Due to the "degree difference" in the coupling between green finance and green technology innovation, when the coupling coordination is low under the influence of poor matching between financial support and technology R&D, it may be difficult for the two to form a collaborative force, and the driving effect on reducing pollution and carbon is weak; When the coupling cooperative scheduling crosses a certain threshold, the synergistic effect may increase exponentially. This setting can directly answer "when the coupling coordination degree reaches what level, the pollution reduction and carbon reduction effect will be significant", reveal the "qualitative change critical point" of the core variable itself, and reflect the internal nonlinear characteristics of the relationship[44].As an external environmental variable, the threshold effect of financial development reflects the "basic condition constraints" of the core relationship. The level of financial development is the "support base" for the coupling of green finance and technological innovation. When the level of financial development is low due to the imperfect financial market, narrow financing channels, lack of risk dispersion mechanism and other factors, even if green finance and technological innovation form a certain coupling, it may be difficult to effectively promote pollution reduction and carbon reduction due to insufficient capital scale, low allocation efficiency and other issues[45]; When the financial development level crosses the threshold after the capital market is mature, the green financial instruments are abundant, and the risk pricing is reasonable, the resource allocation efficiency of the coupling system will be significantly improved, and its pull effect on reducing pollution and carbon will be significantly enhanced[46]. This setting can reveal "how the external financial environment restricts the exertion of the core coupling effect", clarify the "preconditions" for the effectiveness of the core mechanism, and provide more specific boundaries for policy-making. Based on the above analysis, the following hypothesis was proposed:

**Hypothesis 3.** The level of financial development plays a threshold effect in the process of the coupling and coordination of green finance and green technology innovation affecting the synergistic effect of pollution and carbon reduction. When the level of financial development is low, the coupling and coordination of green finance and green technology innovation may hinder the improvement of the synergistic effect. With the improvement of the level of financial development, the synergistic effect of the coupling and coordination of green finance and green technology innovation on pollution and carbon reduction will gradually increase.

*2.4. Spatial spillover effect*

There may be a spatial spillover effect in the collaborative governance of pollution reduction and carbon reduction, and the interaction of carbon emission strategies in different regions may lead to the convergence of carbon emissions in adjacent regions. Due to the fluidity of the atmosphere, air pollution is not restricted by specific areas, resulting in the possible diffusion of pollutants to adjacent areas[47]. Differences in environmental regulatory policies between regions will also trigger enterprises to transfer pollutant emissions to adjacent areas with relatively loose regulation, leading to the phenomenon of "pollution transfer"[48].The coupling and coordination of green finance and green technology innovation may also produce spatial spillover effects. For example, the development of local green finance can drive green investment, green technology innovation and green industry upgrading in adjacent areas[49]; The implementation of the local green policy will form a demonstration effect, which will promote the neighboring areas to imitate or learn from[50]; After the gathering of local green resources, the green transformation of surrounding areas will be driven by the flow of capital, technology and talents[51].Green technology innovation may also have a spatial spillover effect. After a region has made a technological breakthrough in the green and low-carbon field, its innovation achievements will spread to the surrounding and even wider regions through knowledge diffusion, industrial linkage, policy demonstration and environmental spillover, so as to drive the green transformation of adjacent regions[52]. For example, new energy, energy conservation and emission reduction, clean production and other technologies can be exported through enterprise investment, scientific research cooperation, technology trade and other ways to promote the industrial upgrading of surrounding areas; Successful green policies can also be imitated and promoted to form a regional collaborative governance pattern. At the same time, environmental benefits such as air quality improvement and water resources protection are cross regional, making the positive effect of green technology innovation beyond geographical boundaries, promoting regional green coordinated development, and helping to achieve the "double carbon" goal and sustainable development[53]. Based on the above analysis, the following hypothesis was proposed:

**Hypothesis 4.** The synergy effect of reducing pollution and carbon in this region will be affected not only by the coupling coordination relationship between green finance and green technology innovation, but also by the coupling coordination relationship in adjacent regions.

**3. Research Design**
*3.1. Data Source and Processing*
　　This study is based on the panel data of 266 prefecture level cities in China from 2007 to 2023. The original data are mainly from the China Urban Statistical Yearbook, the China Industrial statistical yearbook, the China energy statistical yearbook, the China Environmental Statistical Yearbook, the China Science and technology statistical yearbook, the China Financial Yearbook, the yearbooks of local cities, the guotai'an database and the EPS database platform. The prefecture level cities with excessive data missing will be eliminated. For a small amount of missing data, this paper uses the average growth rate in recent five years and linear interpolation method to standardize the processing by referring to the relevant literature to ensure that the data continuity meets the analysis requirements. The statistical characteristics of the main variables are described in Table 1.

**Table 1.** Descriptive statistics of main variables

| Symbol | Variable Name | Sample | Mean | Std | Min | Max |
|---|---|---|---|---|---|---|
| rpc | Reduction of pollution and carbon emissions | 4522 | 0.1829 | 0.0732 | 0.023 | 0.5085 |
| cogg | The coupling and coordination level of green finance and green technological innovation | 4522 | 0.2532 | 0.0978 | 0.0713 | 0.8241 |
| lngdp | Economic development level | 4522 | 10.6277 | 0.6776 | 8.1309 | 12.4863 |
| urban | Urbanization rate | 4522 | 0.5446 | 0.1602 | 0.1152 | 1.0000 |
| struc | Industrial structure | 4522 | 0.4592 | 0.1109 | 0.116 | 0.8508 |
| tech | Scientific and technological level | 4522 | 0.0168 | 0.017 | -0.0001 | 0.2068 |
| gov | Degree of government intervention | 4522 | 0.1876 | 0.099 | 0.0437 | 1.0268 |

*3.2. Description of Variables*

3.2.1. Explained Variable

Referring to the relevant literature, this paper calculates the pollution reduction system (pollution) and carbon reduction system (carbon) from the three dimensions of sulfur dioxide, smoke (dust) and wastewater, as well as the intensity of carbon dioxide and carbon emissions. The specific indicators are explained in table 2.

**Table 2.** Index system of pollution reduction system and carbon reduction system

| Subsystem | Dimension | Index Interpretation | Symbol |
|---|---|---|---|
| Pollution reduction system | Sulfur dioxide | Industrial sulfur dioxide emission (ton) | — |
| | Smoke (dust) | Industrial smoke (dust) emission (ton) | — |
| | Wastewater | Industrial wastewater discharge (10000 tons) | — |
| Carbon reduction system | Carbon dioxide | Carbon dioxide emissions (10000 tons) | — |
| | Carbon emission intensity | Regional CO2 emissions/regional GDP (10000 tons/100 million yuan) | — |

The coupling degree reflects the strong and weak relationship among subsystems in the system, and the coordination degree reflects whether subsystems promote each other at a high level or restrict each other at a low level. In order to reflect the synergistic relationship between pollution reduction and carbon reduction, the entropy weight TOPSIS is used to calculate the comprehensive index values of the two subsystems, and the coupling coordination degree model is used to calculate the synergistic level of the explained variables[54].

$$C = \frac{2\sqrt{pollution \times carbon}}{pollution + carbon} \quad (1)$$

$$D = \sqrt{C \times T} = \sqrt{C \times (0.5 pollution + 0.5 carbon)} \quad (2)$$

Since the indicators in the two subsystems of pollution reduction and carbon reduction are reverse indicators, the calculated coupling coordination degree is also reverse indicators. The smaller the value, the higher the degree of interdependence of carbon emissions, and the more obvious the synergistic effect of pollution reduction and carbon reduction.

### 3.2.2. Explanatory Variable

Referring to relevant literature[55], the guidance on building a green financial system puts forward that the green financial system mainly includes green credit, green bonds, green stock index and related products, green development fund, green insurance, carbon finance and other financial formats. Based on the above guidance and literature, the green finance subsystem (GreenF) constructed in this paper is constructed from seven dimensions: green credit, green investment, green insurance, green bonds, green payment, green fund and green equity. Referring to relevant research[56,57], this paper measures the green technology innovation and development subsystem (GreenT) from the perspectives of innovation input, innovation output and innovation environment. The specific indicators are explained in Table 3.

**Table 3.** Index system of green finance subsystem and green technology innovation and development subsystem

| Subsystem | Primary indicator | Secondary index | Index Interpretation | Symbol |
|---|---|---|---|---|
| GreenF | Green credit | Proportion of credit for environmental protection projects | Total credit for environmental protection projects/total credit | + |
| | Green investment | Proportion of environmental pollution control investment in GDP | Investment in environmental pollution control/gdp | + |
| | Green insurance | Promotion degree of environmental pollution liability insurance | Environmental pollution liability insurance income/total premium income | + |
| | Green bond | Development degree of green bonds | Total issuance of green bonds/total issuance of all bonds | + |
| | Green payment | Proportion of fiscal expenditure on environmental protection | Financial environmental protection expenditure/financial general budget expenditure | + |
| | Green Fund | Proportion of green funds | Total market value of green funds/total market value of all funds | + |
| | Green equity | Development depth of green rights and interests | Total amount of carbon trading, energy rights trading, emission rights trading/equity market transactions | + |
| GreenT | Innovation investment | Government input | Local fiscal expenditure on science and Technology (10000 yuan) | + |
| | | Scientific researchers | Number of full-time teachers in Colleges and universities (person) | + |
| | | Patent application for | Number of green invention | + |

| | | | |
|---|---|---|---|
| Innovation output | invention | patent applications | |
| | Invention patent authorization | Number of green invention patents authorized | + |
| | Utility model patent application | Number of green utility model patent applications (item) | + |
| | Utility model patent authorization | Number of green utility model patents authorized (items) | + |
| Innovation environment | Public library books | Total collection of books in the public library (1000 volumes, pieces) | + |

In order to reflect the collaborative relationship between green finance and green technology innovation, the entropy weight TOPSIS is used to calculate the comprehensive index values of the two subsystems, and the core explanatory variable, the coupling coordinated development level of green finance and green technology innovation, is calculated according to the two comprehensive index values with reference to relevant literature[58]. The specific results are shown in Table 4. The first column in the table indicates the coordination stage, which in order is extreme imbalance [0,0.1), severe imbalance [0.1,0.2), moderate imbalance [0.2,0.3), mild imbalance [0.3,0.4), verge of imbalance [0.4,0.5), barely coordination [0.5,0.6), primary coordination [0.6,0.7), intermediate coordination [0.7,0.8), good coordination [0.8,0.9), and high-quality coordination[0.9,1]. The first row represents the coupling stage, which in order is low-level coupling [0.0,0.3], run-in stage (0.3,0.5], antagonistic stage (0.5,0.8], and high-level coupling (0.8,1.0].

Table 4. Coupling and coordinating distribution of green finance and green technology innovation in China's cities

| Coordination phase (d) | Coupling phase (c) | | | | Total |
|---|---|---|---|---|---|
| | Low-level coupling | Run-in stage | Antagonistic stage | High-level coupling | |
| Extreme imbalance | 4 | 19 | 5 | 0 | 28 |
| Severe imbalance | 706 | 518 | 163 | 32 | 1419 |
| Moderate imbalance | 229 | 1098 | 619 | 68 | 2014 |
| Mild imbalance | 2 | 84 | 426 | 133 | 645 |
| Verge of imbalance | 0 | 1 | 58 | 231 | 290 |
| Barely coordination | 0 | 0 | 3 | 88 | 91 |
| Primary coordination | 0 | 0 | 0 | 29 | 29 |
| Intermediate coordination | 0 | 0 | 0 | 5 | 5 |
| Good coordination | 0 | 0 | 0 | 1 | 1 |
| High-quality coordination | 0 | 0 | 0 | 0 | 0 |
| Total | 941 | 1720 | 1274 | 587 | 4522 |

3.2.3. Mediating variable

Energy efficiency is selected as the intermediary variable. Energy efficiency (lnee) is measured in logarithms of GDP per unit of regional energy consumption. As for GDP per unit energy consumption, the provincial energy consumption data is decomposed into prefecture level cities according to the lighting data value by using the linear model without intercept[59]. Then, divide the GDP of each city by the decomposed energy consumption of each city to obtain the GDP per unit energy consumption of each region.

3.2.4. Threshold variable

In order to reveal the possible nonlinear impact mechanism of the coupling between green finance and green technology innovation and the level of financial development on pollution reduction and carbon reduction, the core explanatory variable coupling between green finance and green technology innovation itself is selected as the threshold variable, while the level of financial development is selected as the threshold variable. As an external environmental variable, the threshold effect of financial development reflects the "basic condition constraints" of the core relationship. The level of financial development (fdl) is measured logarithmically by the sum of the balance of loans of local financial institutions at the end of the year and the balance of deposits of financial institutions at the end of the year as a proportion of the regional GDP.

3.2.5. Control Variable

Drawing on previous relevant studies, the following control variables are introduced, specifically including: the level of economic development (lngdp), and selecting the logarithm of regional per capita GDP to measure the index; The urbanization rate (urban) is measured by the proportion of non-agricultural population in the registered population; Industrial structure (struc) is measured by the proportion of the added value of the regional secondary industry in the regional GDP; The level of science and Technology (tech) is measured by the proportion of regional science expenditure in the general budget of local finance; The degree of government intervention (gov) is measured by the proportion of the expenditure of 10000 yuan in the general budget of local finance in the GDP of the region.

*3.3. Model setting*

3.3.1. The Baseline Model

Based on the above analysis, in order to investigate the direct impact of green finance and green technology innovation on pollution and carbon reduction, this paper constructs the following double fixed effect model:

$$rpc_{it} = \alpha_0 + \alpha_1 cogg_{it} + \alpha_2 control_{it} + year_t + city_i + \varepsilon_{it} \qquad (3)$$

Where, $rpc_{it}$ is the synergy level of the explained variable in reducing pollution and carbon, $cogg_{it}$ is the core explanatory variable coupling green finance and green technology innovation, $control_{it}$ is the control variable, $year_t$ is the fixed effect of the year, $city_i$ is the fixed effect of the city, and $\varepsilon_{it}$ is the random error term. Where subscripts $i$ and $t$ represent prefecture level cities and years respectively.

3.3.2. Mediating effect model

In order to investigate the indirect impact mechanism of green finance and green technology innovation on pollution and carbon reduction, the paper tests whether the dual green coupling promotes pollution and carbon reduction by improving energy efficiency. Using the relevant ideas for reference[60], this paper constructs the mediation effect test model as follows:

$$rpc_{it} = \alpha_0 + \alpha_1 cogg_{it} + \alpha_2 control_{it} + year_t + city_i + \varepsilon_{it} \qquad (4)$$
$$\ln e\,e_{it} = \beta_0 + \beta_1 cogg_{it} + \beta_2 control_{it} + year_t + city_i + \varepsilon_{it} \qquad (5)$$
$$rpc_{it} = \gamma_0 + \gamma_1 \ln e\,e_{it} + \gamma_2 cogg_{it} + \gamma_3 control_{it} + year_t + city_i + \varepsilon_{it} \qquad (6)$$

Among them, $\ln e\,e_{it}$ is the intermediary variable energy efficiency, and the other variables are consistent with the above.

3.3.3. Panel threshold model

When the threshold variable is the core explanatory variable, green finance and green technology innovation are coupled, the model is set as follows:

$$rpc_{it} = C + \delta_1 cogg_{it} \times I \cdot (cogg_{it} \leq \theta_1) + \delta_2 cogg_{it} \times I \cdot (\theta_1 < cogg_{it} \leq \theta_2)$$
$$+ \cdots + \delta_{n+1} cogg_{it} \times I \cdot (\theta_n < cogg_{it} \leq \theta_{n+1}) + \delta_0 control_{it} + city_i + \varepsilon_{it} \quad (7)$$

When the threshold variable is the level of financial development, the model is set as follows:

$$rpc_{it} = C + \delta_1 cogg_{it} \times I \cdot (fdl_{it} \leq \theta_1) + \delta_2 cogg_{it} \times I \cdot (\theta_1 < fdl_{it} \leq \theta_2)$$
$$+ \cdots + \delta_{n+1} cogg_{it} \times I \cdot (\theta_n < fdl_{it} \leq \theta_{n+1}) + \delta_0 control_{it} + city_i + \varepsilon_{it} \quad (8)$$

Where $cogg_{it}$ is the core explanatory variable and threshold variable, $fdl_{it}$ is the threshold variable, $\theta_1, \theta_2, \ldots, \theta_n$ is $n$ threshold values, $\delta_1, \delta_2, \ldots, \delta_n$ is the regression coefficient of different threshold interval, $\delta_0$ is the coefficient of control variable, and the other variables are the same as above.

## 4. Results and Discussion

### 4.1. Baseline Results

This paper uses the panel data of prefecture level cities to study the impact of the coupling of green finance and green technology innovation on the synergistic effect of pollution and carbon reduction. The specific regression results are shown in Table 5. Among them, column (1) only considers the fixed effects of city and time, and columns (2) to (7) gradually add control variables. According to table 5, the coefficient of the core explanatory variable $cogg$ is significantly negative whether the control variable is considered or not. It can be seen from column (6) that on the basis of controlling individual effects, time effects and control variables, the impact of the coupling of green finance and green technology innovation on the synergistic effect of pollution and carbon reduction is significantly negative at the level of 1%, indicating that the coupling of green finance and green technology innovation has significantly promoted the city's pollution and carbon reduction, H1 has been verified.

Table 5. Benchmark regression results

| Variables | (1) rpc | (2) rpc | (3) rpc | (4) rpc | (5) rpc | (6) rpc |
|---|---|---|---|---|---|---|
| cogg | -0.133*** | -0.131*** | -0.130*** | -0.128*** | -0.103*** | -0.095*** |
|  | (-6.35) | (-6.36) | (-6.31) | (-6.21) | (-4.71) | (-4.32) |
| lngdp |  | 0.036*** | 0.035*** | 0.036*** | 0.038*** | 0.043*** |
|  |  | (12.58) | (12.13) | (10.92) | (11.33) | (11.91) |
| urban |  |  | 0.023*** | 0.022*** | 0.022*** | 0.020** |
|  |  |  | (2.70) | (2.67) | (2.60) | (2.33) |
| struc |  |  |  | -0.011 | -0.013 | -0.011 |
|  |  |  |  | (-0.86) | (-1.07) | (-0.86) |
| tech |  |  |  |  | -0.191*** | -0.177*** |
|  |  |  |  |  | (-3.38) | (-3.14) |
| gov |  |  |  |  |  | 0.053*** |
|  |  |  |  |  |  | (3.65) |
| _cons | 0.217*** | -0.167*** | -0.169*** | -0.180*** | -0.200*** | -0.258*** |
|  | (40.61) | (-5.40) | (-5.47) | (-5.39) | (-5.91) | (-6.91) |
| city&year | YES | YES | YES | YES | YES | YES |
| N | 4522 | 4522 | 4522 | 4522 | 4522 | 4522 |
| $R^2$ | 0.818 | 0.825 | 0.825 | 0.825 | 0.826 | 0.826 |

### 4.2. Endogenous treatment and robustness test

#### 4.2.1. Endogenous treatment

In order to solve the endogenous problem caused by missing variables and two-way causality, referring to relevant research[61], the first stage of coupling lag between green finance and green technology innovation is selected as the instrumental variable for 2SLS regression, and the results are shown in columns (1) and (2) of table 6. From the first stage regression, it can be seen that the selected instrumental variables are effective. The second stage regression results show that the promotion effect of the coupling of green finance and green technology innovation on reducing pollution and carbon is basically consistent with the above results. At the same time, LM statistics reached 39.402, P value was 0.000, Cragg-Donald Wald F statistics and Kleibergen-Paap rk Wald F statistics were 1673.254 and 225.684, respectively, indicating that there was no problem of insufficient identification and weak instrumental variables, and the correlation characteristics were met. The coupling of green finance and green technology innovation still showed a significant inhibitory effect on urban pollution and carbon reduction. In order to solve the possible problems of autocorrelation and heteroscedasticity, the System GMM method is further used to re regress the benchmark model to verify the robustness of the model test results. From the results of model (3) in Table 6, it is found that the P values of AR (1), AR (2) and Hansen are 0.000, 0.122 and 0.588, respectively. The impact coefficient of the coupling of green finance and green technology innovation on pollution and carbon reduction is still significantly negative and has not changed significantly.

**Table 6.** Endogenous treatment results

| Variables | (1) First (cogg) | (2) Second (rpc) | (3) Sys-GMM (rpc) |
|---|---|---|---|
| cogg | | -0.216** | -0.072*** |
| | | (-2.43) | (-5.63) |
| L.rpc | | | 1.075*** |
| | | | (81.14) |
| L.cogg | 0.542*** | | |
| | (15.02) | | |
| Kleibergen-Paap rk LM | 39.402(0.000) | | |
| Cragg-Donald Wald F | 1673.254(＞16.380) | | |
| Kleibergen-Paap rk Wald F | 225.684(＞16.380) | | |
| AR(1) P | | | 0.000 |
| AR(2) P | | | 0.122 |
| Hansen P | | | 0.588 |
| _cons | 0.212*** | -0.260*** | 0.136** |
| | (7.14) | (-2.31) | (2.69) |
| control | YES | YES | YES |
| city&year | YES | YES | YES |
| N | 4,256 | 4,256 | 4256 |
| $R^2$ | 0.297 | 0.364 | |

Note: t value in brackets, * P<0.1, ** P<0.05, *** P<0.01.

### 4.2.2. Robustness Test

In order to test the reliability of the above research results, the following methods are used for robustness test. The first is to replace the explained variables, normalize the range of the three indicators through the three indicators of sulfur dioxide emissions, smoke (dust) emissions and wastewater emissions[62], and construct the comprehensive index of environmental pollution index by using the improved entropy weight TOPSIS method, and then use the logarithm of the cross product of carbon emissions and environmental pollution indicators to represent the level of pollution reduction and carbon reduction. The second is to increase the control variables, and select three control variables: financial investment intensity (fii), human capital level (lhc) and degree of openness (dft), in order to control other influencing factors as much as possible, reduce the deviation caused by missing variables, and make the estimation results of core explanatory variables more accurately reflect the true relationship between them and the explained variables. The third is to change the clustering standard and raise the clustering standard to provincial level, which is to examine the data from a more macro level, change the aggregation mode of the data, and thus change the correlation structure of the data. Results as shown in models (1) to (3) in Table 7, the impact coefficient of the coupling of green finance and green technology innovation on pollution and carbon reduction is still significantly negative, indicating that the result is robust.

**Table 7.** Robustness test results

| Variables | (1) Replace explained variable | (2) Add control variable | (3) Change clustering criteria |
|---|---|---|---|
| cogg | -1.428*** | -0.065*** | -0.188** |
|  | (-2.90) | (-2.92) | (-2.00) |
| fii |  | 0.001*** |  |
|  |  | (3.55) |  |
| lhc |  | -0.415*** |  |
|  |  | (-4.98) |  |
| dft |  | 0.019*** |  |
|  |  | (4.40) |  |
| _cons | 3.110*** | -0.212*** | -0.368 |
|  | (3.72) | (-5.60) | (-1.39) |
| control | YES | YES | YES |
| city&year | YES | YES | YES |
| $N$ | 4522 | 4522 | 442 |
| $R^2$ | 0.849 | 0.828 | 0.903 |

Note: t value in brackets, * $P<0.1$, ** $P<0.05$, *** $P<0.01$.

### 4.3. Heterogeneity analysis

Further analyze whether there are regional differences in the impact of the coupling of green finance and green technology innovation on the synergistic effect of pollution reduction and carbon reduction. Taking into account the heterogeneity of cities, the samples are divided according to the regional location of cities, resource endowment and the development level of the Yangtze River Basin, so as to further analyze the heterogeneous effect of the coupling of green finance and green technology innovation on the synergistic effect of pollution reduction and carbon reduction.

4.3.1. Regional heterogeneity

It is divided into three regions according to regions. The eastern, central and western regions are tested. The regional division standard comes from the "Seventh Five Year Plan" adopted at the fourth session of the Sixth National People's Congress in 1986. The results by region are shown in columns (1) - (3) of table 8. The impact of the coupling of green finance and green technology innovation on pollution and carbon reduction in the eastern and western regions is negative through the 1% significance level. Referring to the previous benchmark regression conclusion, it shows that green finance and green technology innovation promote the synergy of pollution and carbon reduction in the eastern and western regions. The regression coefficient of double green coupling on reducing pollution and carbon in the central region is positive and not significant, which reflects the obvious regional differentiation of the coupling effect of green finance and technological innovation. The possible reason is that the economy in the East is developed, the foundation of green finance and green technology innovation is good, and the coupling effect has been deeply integrated into the practice of reducing pollution and carbon. The East has perfect market mechanism and technology transformation ability. The combination of green finance and green technology innovation can effectively promote pollution reduction and carbon peak, and the synergistic effect is significant. However, the ecological environment in Western China is fragile and the demand for reducing pollution and carbon is urgent. The "forced back effect" of the coupling of green finance and green technology innovation is more significant.

4.3.2. Resource heterogeneity

According to the national sustainable development plan for resource-based cities (2013-2020), the research sample is divided into resource-based cities and non resource-based cities. As shown in columns (4) and (5) of Table 8, the coupling of green finance and green technology innovation in non resource cities has a significant negative impact on pollution reduction and carbon reduction at the 1% level, while the regression of resource-based cities is not significant. This indicates that the dual green coupling can promote pollution reduction and carbon reduction in non resource cities. The reason may be that non resource-based cities have no path dependence on traditional resource industries, the market mechanism is more flexible, green finance and green technology innovation can be rapidly integrated, and the synergy effect is significant. In addition, non resource-based cities have diversified industries, and there is a large synergy between green finance and green technology innovation. Financial resources are more likely to flow to green industries. Green technology innovation and the demand for reducing pollution and carbon can be directly matched from the aspects of industrial energy-saving technology, which can significantly promote the synergy of reducing pollution and carbon, and the coupling effect can be effectively released.

4.3.3. Heterogeneity in the Yangtze River Basin

According to the development planning outline of the Yangtze River economic belt, the research samples are divided into cities along the Yangtze River Basin and non cities along the Yangtze River Basin. The results of dividing cities along the Yangtze River Basin are shown in columns (6) and (7) of Table 8. The impact of the coupling of green finance and green technology innovation on the reduction of pollution and carbon in cities along the non Yangtze River Basin is significantly negative at the 1% level, while the regression of cities along the Yangtze River Basin is not significant. The above shows that the promotion effect of double green coupling on the reduction of pollution and carbon in cities along the non Yangtze River Basin is more significant than that in cities along the Yangtze River Basin. The possible reason is that the industrial layout of cities in non Yangtze River Basin is more flexible, and there is no additional constraint of basin ecological protection. Green finance and green technology innovation can be quickly transformed into synergy benefits. There is no strong policy constraint of "great protection of the Yangtze River" in the non Yangtze River Basin. The "market orientation" of green finance and green technology innovation is more obvious. Financial resources can be flexibly allocated to support new energy, intelligent manufacturing and other aspects. Green technology innovation and pollution reduction and carbon reduction can be directly matched in industrial energy conservation, VOCs governance and other aspects. The coupling effect has been effectively released, and the effect intensity is equivalent to that in the eastern region.

Table 8. Heterogeneity test results

| Variables | (1) rpc | (2) rpc | (3) rpc | (4) rpc | (5) rpc | (6) rpc | (7) rpc |
|---|---|---|---|---|---|---|---|
| cogg | -0.176*** | 0.016 | -0.344*** | -0.009 | -0.155*** | 0.044 | -0.176*** |
|  | (-5.20) | (0.42) | (-4.73) | (-0.20) | (-5.87) | (1.50) | (-5.59) |
| _cons | -0.430*** | -0.296*** | -0.069 | -0.340*** | -0.156*** | -0.478*** | -0.172*** |
|  | (-6.70) | (-5.29) | (-0.68) | (-6.20) | (-3.02) | (-7.62) | (-3.27) |
| control | YES | YES | YES | YES | YES | YES | YES |
| city&year | YES | YES | YES | YES | YES | YES | YES |
| N | 1598 | 2023 | 901 | 1853 | 2669 | 1751 | 2771 |
| $R^2$ | 0.840 | 0.814 | 0.807 | 0.817 | 0.837 | 0.852 | 0.809 |

Note: t value in brackets, * P<0.1, ** P<0.05, *** P<0.01. Columns (1) - (3) are the eastern, central and western regions in turn, columns (4) and (5) are resource-based cities and non resource-based cities in turn, and columns (6) and (7) are cities along the Yangtze River Basin and non Yangtze River Basin cities in turn.

*4.4. Mediating effect Test*

The results shown in Table 8, the regression coefficient $\alpha_1$ (-0.095) is significantly negative at the 1% significance level through column (1), indicating that the synergistic effect of pollution and carbon reduction will be affected by the coupling of green finance and green technology innovation, and further mechanism analysis. In column (2), the regression coefficient $\beta_1$ (0.674) is significantly positive at the level of 1%, indicating that the coupling of green finance and green technology innovation can significantly improve energy efficiency; The product of the regression coefficient $\beta_1$ (0.674) of column (2) and the influence coefficient $\gamma_1$ (-0.029) of energy efficiency on pollution reduction and carbon reduction of column (3) is the same sign as the regression coefficient $\gamma_2$ (-0.076) of column (3), indicating that there is a mediating effect; Comparing the coefficient $\alpha_1$ (-0.095) in the benchmark regression of column (1) with the regression coefficient $\gamma_2$ (-0.076) of column (3), it is found that the absolute value of $\gamma_2$ is lower than that of $\alpha_1$, and it is still significantly negative at the level of 1%, indicating that the mediation effect is partial mediation effect. Causal mediating effect analysis shows that the direct and indirect impact of the coupling of green finance and green technology innovation on the synergy of urban pollution and carbon reduction is very significant, and 20.3% of it comes from the intermediary transmission mechanism of energy efficiency. In addition, the results of Sobel test further show that the coupling of green finance and green technology innovation can significantly improve energy efficiency and thus promote urban pollution and carbon reduction, which verifies H2.

Table 9. Mediating effect of energy efficiency

| Variables | (1) rcp | (2) lnee | (3) rcp | | |
|---|---|---|---|---|---|
| cogg | -0.095*** | 0.674*** | -0.076*** | Indirect effect | -0.019*** |
|  | (-4.32) | (5.53) | (-3.47) | Direct effect | -0.076*** |
| lnee |  |  | -0.029*** | Total effect | -0.095*** |
|  |  |  | (-10.44) | PTE | 0.203 |
| _cons | -0.258*** | 4.983*** | -0.116*** | | |
|  | (-6.91) | (24.10) | (-2.94) | | |
| Sobel test |  | -0.019*** |  | | |
|  |  | (0.004) |  | | |
| control | YES | YES | YES | | |
| city&year | YES | YES | YES | | |
| $N$ | 4522 | 4522 | 4522 | | |
| $R^2$ | 0.826 | 0.925 | 0.831 | | |

Note: In Sobel test, robust standard error is shown in brackets, and t value is shown in other brackets, * P<0.1, ** P<0.05, *** P<0.01.

## 5. Further analysis

### 5.1. Threshold effect

This study first follows the research paradigm of previous scholars[63], and sets the threshold parameters and specific thresholds in the threshold regression model. The test results are shown in Table 10.

Table 10. Bootstrap sampling results

| Variables | Threshold quantity | F | P | Bootstrap | Critical value | | |
|---|---|---|---|---|---|---|---|
|  |  |  |  |  | 10% | 5% | 1% |
| cogg | Single | 105.47 | 0.0000 | 500 | 33.2117 | 39.1390 | 56.2683 |
|  | Double | 27.97 | 0.0960 | 500 | 27.0017 | 54.8898 | 111.8904 |
|  | Triple | 11.94 | 0.6520 | 500 | 38.9293 | 54.3335 | 102.7663 |
| fdl | Single | 157.95 | 0.0000 | 500 | 37.4549 | 43.4557 | 57.3023 |
|  | Double | 38.08 | 0.0640 | 500 | 33.7897 | 41.8194 | 55.2521 |
|  | Triple | 32.91 | 0.6460 | 500 | 69.0888 | 81.6137 | 104.7493 |

As demonstrated in Table 10, the threshold variable—namely, the coupling between green finance and green-technology innovation—passes both the single- and double-threshold tests at the 10 % significance level, whereas the triple-threshold test is not significant, indicating the presence of a double threshold. Likewise, the threshold variable financial-development level passes the single- and double-threshold tests at the 10 % significance level but fails the triple-threshold test, likewise evidencing a double-threshold effect.

The threshold regression results are shown in Table 11. When the threshold variable is the level of financial development, and the threshold value is lower than or equal to 0.5080, the coupling coefficient of green finance and green technology innovation is -0.155, which is significant at the 1% level, indicating that in the low financial development range, the coupling of green finance and green technology innovation can still play a certain role in promoting the coordinated development of pollution and carbon reduction. This may be because the regions with low financial development are easier to receive policy guidance and the urgent need for technological innovation support makes it more effective. When the financial development level is between 0.5080 and 1.1680, the coefficient of coupling of green finance and green technology innovation becomes -0.198, which is significant at the 1% level, indicating that after the financial development level rises to this range, the promotion effect of the coupling of green finance and green technology innovation on reducing pollution and carbon is more obvious. At this stage, it may be that after the improvement of the level of financial development, the integration with regional green finance has been accelerated, and green technology innovation has been funded, thus optimizing the allocation of resources and reducing unnecessary emissions. When the financial development level is higher than 1.1680, the coupling of green finance and green technology innovation coefficient is further increased to -0.252, which is significant at the 1% level, indicating that with the further improvement of the financial development level, the marginal emission reduction effect of the coupling of green finance and green technology innovation continues to increase. When the threshold variable is the coupling of green finance and green technology innovation, the conclusion is the same. The development of green finance and green technology innovation has led to the rapid agglomeration of capital, talent and technology elements. This policy guidance, capital inflow and the concentration of knowledge intensive resources can create development opportunities for high-tech enterprises. To sum up, the level of financial development plays a regulatory role between the coupling of green finance and green technology innovation and the coordinated development of pollution reduction and carbon reduction. When the level of financial development crosses the key threshold, the green benefits of the coupling of green finance and green technology innovation can be brought into full play. H3 was verified.

**Table 11.** Threshold regression results

| Variables | (1) rpc | (2) rpc |
|---|---|---|
| cogg(cogg≤0.3795) | -0.145*** | |
| | (-3.01) | |
| cogg(0.3795＜cogg≤0.4374) | -0.176*** | |
| | (-3.69) | |
| cogg(cogg＞0.4374) | -0.208*** | |
| | (-4.00) | |
| cogg(fdl≤0.5080) | | -0.155*** |
| | | (-2.83) |
| cogg(0.5080＜fdl≤1.1680) | | -0.198*** |
| | | (-3.65) |
| cogg(fdl＞1.1680) | | -0.252*** |
| | | (-4.76) |
| _cons | 0.253*** | 0.242*** |
| | (7.23) | (6.81) |
| Control | YES | YES |
| city | YES | YES |
| $N$ | 4522 | 4522 |

|  | $R^2$ | 0.305 | 0.318 |

Note: t value in brackets, * P<0.1, ** P<0.05, *** P<0.01.

*5.2. Analysis of spatial spillover effect*

In order to further explore whether the coupling of green finance and green technology innovation has a spatial spillover effect on the synergistic effect of pollution reduction and carbon reduction, the following spatial effect model is constructed:

$$rpc_{it} = \alpha_0 + \rho W rpc_{it} + \alpha_1 cogg_{it} + \alpha_2 control_{it} + \gamma_1 W cogg_{it} + \gamma_2 W control_{it} + year_i + city_i + \varepsilon_{it} \quad (9)$$

Where $W$ is the spatial weight matrix and $\rho$ is the spatial autocorrelation coefficient. In order to ensure the robustness of the results, in addition to the adjacent weight matrix, the economic distance weight matrix is added for regression analysis.

Spatial autocorrelation is an important prerequisite for spatial econometric analysis. Based on the adjacency weight matrix, Moran's I was used for spatial autocorrelation test. The results show that Moran's I of the coupling of green finance and green technology innovation and the reduction of pollution and carbon are significantly positive from 2007 to 2023, indicating that there is an obvious spatial autocorrelation between the coupling of green finance and green technology innovation and the reduction of pollution and carbon in all cities during the investigation period, and the two show the distribution characteristics of spatial agglomeration, which is reasonable to use spatial econometric analysis. For the selection of spatial econometric model, LM Test, Hausman test, LR test and Wald test were carried out in turn. Specifically, the LM-error test under the geographical contiguity matrix is insignificant, indicating that the spatial lag model (SLM) should be employed; whereas the LM-lag test under the economic-distance matrix is insignificant, suggesting that the spatial error model (SEM) is appropriate. Upon further inspection of the LR and Wald test results, both decisively reject the null hypotheses of degenerating to either the spatial lag or the spatial error model, implying that the more general spatial Durbin model (SDM) is preferable[64]. Consequently, the panel SDM with individual fixed effects was selected for the final regression analysis. The test results are shown in Table 12.

Table 12. Verification of spatial econometric model

| Test | Geographic adjacency matrix | | Economic distance matrix | |
| --- | --- | --- | --- | --- |
|  | Statistic | P | Statistic | P |
| LM-lag | 933.4500 | 0.0000 | 8.1230 | 0.0040 |
| Robust LM-lag | 732.7230 | 0.0000 | 0.9190 | 0.3380 |
| LM-error | 201.7700 | 0.0000 | 17.9320 | 0.0000 |
| Robust LM-error | 1.0430 | 0.3070 | 10.7280 | 0.0010 |
| Hausman | 75.24 | 0.0000 | 497.39 | 0.0000 |
| LR(SDM&SAR) | 36.00 | 0.0000 | 29.43 | 0.0001 |
| LR(SDM&SEM) | 28.17 | 0.0000 | 35.53 | 0.0000 |
| Wald(SDM&SAR) | 36.22 | 0.0000 | 29.49 | 0.0000 |
| Wald(SDM&SEM) | 28.27 | 0.0000 | 35.55 | 0.0000 |

It can be seen from Table 13 that the spatial autoregressive coefficient ρ of pollution reduction and carbon reduction under both weight matrices is significantly positive at the level of 1%, indicating that there is a positive spatial spillover effect in urban pollution reduction and carbon reduction, that is, the improvement of pollution reduction and carbon reduction effect in adjacent cities will aggravate the pollution reduction and carbon reduction effect in local cities.

The coupling of green finance and green technology innovation (Main) and its spatial interaction (Wx) coefficient are significantly negative, indicating that the coupling of green finance and green technology innovation has a significant negative spatial spillover effect on urban pollution and carbon reduction, that is, the coupling of green finance and green technology innovation can not only effectively reduce the pollution and carbon reduction in local cities, but also reduce the pollution and carbon reduction in adjacent cities. Further, in order to further study the marginal effect of the coupling of green finance and green technology innovation on urban pollution and carbon reduction, the total spatial effect is divided into direct effect and indirect effect. The results show that: under two different weight matrices, the direct effect, indirect effect and total effect of the coupling of green finance and green technology innovation on urban pollution and carbon reduction are significantly negative, which proves again that the coupling development of green finance and green technology innovation can not only reduce the pollution and carbon reduction of local cities, but also effectively reduce the pollution and carbon reduction of adjacent cities, showing a negative spatial spillover effect. Moreover, the regression results under the two weight matrices are consistent, indicating that the regression results of spatial Dobbin model are robust. H4 was verified.

Table 13. Spatial Dubin regression results

| Variables | Geographic adjacency matrix | | Economic distance matrix | |
|---|---|---|---|---|
| cogg(Main) | -0.152*** | | -0.101*** | |
|  | (0.0238) | | (0.0218) | |
| cogg(Wx) | -0.108*** | | -0.310*** | |
|  | (0.0333) | | (0.0499) | |
| rho | 0.234*** | | 0.275*** | |
|  | (0.0181) | | (0.0255) | |
| sigma2_e | 0.000967*** | | 0.000953*** | |
|  | (2.05e-05) | | (2.01e-05) | |
| Indirect effect | -0.159***(0.0238) | | -0.114***(0.0221) | |
| Direct effect | -0.179***(0.0379) | | -0.452***(0.0645) | |
| Total effect | -0.339***(0.0380) | | -0.566***(0.0665) | |
| control | YES | YES | YES | YES |
| city | YES | YES | YES | YES |
| Observations | 4,522 | 4,522 | 4,522 | 4,522 |
| $R^2$ | 0.042 | 0.042 | 0.020 | 0.020 |

Note: t value in brackets, * P<0.1, ** P<0.05, *** P<0.01.

## 6. Conclusions and Policy Implications

### 6.1. Conclusion

Based on the data of 266 prefecture level cities in China from 2007 to 2023, this paper tests the effect and mechanism of the coupling of green finance and green technology innovation on the synergistic effect of pollution and carbon reduction from multiple dimensions. The results show that: (1) the coupling of green finance and green technology innovation can significantly reduce pollution and carbon reduction. After a series of robustness tests, the conclusion is still valid. (2) The coupling of green finance and green technology innovation can reduce the synergistic effect of urban pollution and carbon reduction by improving energy efficiency. (3) The coupling of green finance and green technology innovation can significantly reduce the collaborative level of pollution reduction and carbon reduction in the eastern and western regions, non resource-based cities and cities along the Yangtze River Basin, but can not significantly inhibit the collaborative level of pollution reduction and carbon reduction in the central region, resource-based cities and cities along the Yangtze River Basin. (4) The inhibitory effect of the coupling of green finance and green technology innovation on the synergistic effect of urban pollution reduction and carbon reduction will be affected by the coupling of green finance and green technology innovation itself and the level of financial development, showing nonlinear characteristics. (5) The coupling of green finance and green technology innovation has a significant negative spatial spillover effect on the synergistic effect of pollution reduction and carbon reduction. The coupling of green finance and green technology innovation can effectively inhibit the collaborative level of pollution reduction and carbon reduction in local cities and adjacent cities.

*6.2.Policy Implications*

First, we should systematically build a multi-level and diversified green financial market, expand the coverage and delivery efficiency of financial instruments such as green credit, green bonds and green insurance, and focus on supporting technology research and development and large-scale application in the field of energy efficiency improvement. The policy design should focus on guiding financial institutions to establish and improve the environmental risk assessment mechanism and green project identification standards, promote the precise investment of green funds into green technology innovation projects with high technology maturity and high emission reduction potential, and give full play to the intermediary transmission function of energy efficiency between green finance and the synergy of pollution reduction and carbon reduction.

Second, formulate differentiated policies based on regional resource endowment, industrial structure and development stage. In the eastern region, we can further deepen the integration of green financial innovation and technology market, and promote the industrialization of high-end green technology; In the central region, efforts should be made to promote the green transformation of traditional industries, strengthen the construction of green technology introduction and absorption capacity, and make up for the green financing gap; In the western region, we should combine ecological compensation with the national clean energy strategy, increase financial transfer payments and policy financial support, and improve the localization application ability of green technology. For resource-based cities, it is necessary to design special transformation financial instruments to help them break through path dependence and reduce the cost of green transformation.

Third, focus on improving the synergy level of green finance and green technology innovation, and adopt the support strategy of phased and dynamic adjustment. In areas with low coupling degree, green technology R&D incubation and green financial market infrastructure should be strengthened; After reaching a certain threshold, the incentive mechanism should be introduced in time to promote technology diffusion and market formation; In the highly collaborative stage, we should focus on building a green science and technology innovation ecosystem and the world's leading green financial infrastructure to achieve continuous enhancement of emission reduction effects. At the same time, we should improve the overall level of financial development, improve the green financial regulatory framework, information disclosure and performance evaluation system, and provide institutional guarantee for achieving a higher level of coupling.

Fourth, break through the restrictions of administrative divisions and build a cross regional collaborative governance system of green finance and technology. Promote the establishment of regional green project databases and environmental information sharing platforms, and encourage the cross regional flow and cooperation of green financial resources. Explore the establishment of ecological compensation and collaborative emission reduction mechanisms in key river basins (such as the Yangtze River and yellow river basins) and urban agglomerations, coordinate environmental standards and green financial policies, and avoid "policy depression" and pollution transfer. Give full play to the radiation and driving role of high coupling regions, and promote green knowledge spillover and overall environmental performance through technical cooperation, talent exchange and capital linkage.

**Author Contributions**
Conceptualization, R.X. (Ruijun Xie) and Q.L.; methodology, software, Q.L.; writing一original draft preparation, R.X. (Ruijun Xie); writing一review and editing. All authors have read and agreed to the published version of the manuscript.
**Funding**
This research was funded by Key Natural Science Research Project of Anhui Provincial Department of Education(2024)"Numerical Methods and Their Applications in Option Pricing Based on Physics-Informed Neural Networks"(Grant No. 2024AH050013).
**Data Availability Statement**
  The data presented in this study are available upon request from the corresponding author.
**Conflicts of Interest**
  The authors declare that they have no known competing financial interests or personal relationships that could have appeared to influence the work reported in this paper.